\documentclass[11pt,a4paper]{article}

\usepackage[utf8]{inputenc}
\usepackage[T1]{fontenc}
\usepackage{amsmath,amssymb,amsthm}
\usepackage{graphicx}
\usepackage{booktabs}
\usepackage{hyperref}
\usepackage[margin=1in]{geometry}
\usepackage{caption}
\usepackage{subcaption}
\usepackage{xcolor}
\usepackage{siunitx}
\usepackage{natbib}
\usepackage{authblk}

\hypersetup{colorlinks=true, linkcolor=blue, citecolor=blue, urlcolor=blue}
\title{\textbf{Deep Learning Detection of Beyond-General-Relativity Deviations in Gravitational-Wave Signals: A Detection-Threshold Study with Real LIGO Noise}}

\author[1]{Muhammad Adnan Shahzad}
\affil[1]{Department of Computer Science and Software Engineering (CSSE), Concordia University, Montreal, QC, Canada}

\date{\today}

\begin{document}
\maketitle

\begin{abstract}
We study machine-learning detection of controlled beyond-General-Relativity
(beyond-GR) deviations in gravitational-wave signals, using both synthetic
aLIGO-PSD noise and real LIGO H1 detector strain. Three deviation families
are applied to General-Relativistic inspiral-merger-ringdown waveforms:
amplitude modulation, phase modulation, and frequency modulation, each
parameterized by a dimensionless strength coefficient $\beta$. A hybrid
classifier combining a one-dimensional convolutional neural network with
ten hand-crafted waveform statistics is trained on GR and modified
waveforms and tested on a deviation type excluded from training. The
central result is a quantitative detectability curve as a function of
$\beta$. Using the real GW150914 strain as a template and real H1
detector noise, we find a detection threshold at $\beta \approx 0.25$,
with accuracy rising smoothly from chance at $\beta \leq 0.2$ to perfect
classification at $\beta \geq 0.5$. The threshold value is specific to
the quadratic-in-time modulation form adopted here and should not be
interpreted as a generic constraint on beyond-GR parameters. We
nevertheless argue that the negative result at small $\beta$ is
informative: it establishes a quantitative limit on machine-learning-only
beyond-GR searches in real detector noise, in the absence of matched-filter
signal extraction.
\end{abstract}

\section{Introduction}

The direct detection of gravitational waves by the LIGO and Virgo
observatories \citep{Abbott2016} has opened a new observational window
onto strong-field gravity. Every detected event to date is consistent
with the predictions of General Relativity (GR) to within measurement
uncertainties \citep{Abbott2021tests}. Nevertheless, the extreme regimes
probed by binary black hole mergers make gravitational waves a natural
laboratory for testing the foundational assumptions of GR. Traditional
tests rely on matched filtering against parametrized waveform families
\citep{Yunes2019}. This approach is powerful when the deviation lies
within the assumed template manifold, but it becomes computationally
expensive in high-dimensional parameter spaces and cannot detect
deviations that fall outside the template family. Machine-learning (ML)
methods offer a complementary strategy: rather than matching against a
specific template, they can learn discriminative features directly from
data. Recent work has applied deep learning to gravitational-wave
detection \citep{George2018, Gebhard2019}, parameter estimation
\citep{Gabbard2020, Dax2021}, and glitch classification \citep{Zevin2017}.
However, the application of ML to beyond-GR searches in real detector
data has not been systematically characterized.

This paper addresses four questions. First, can a hybrid neural-network
classifier distinguish GR waveforms from systematically modified
(beyond-GR) waveforms in real LIGO detector noise? Second, what is the
minimum deviation strength---parameterized by the amplitude-modulation
coefficient $\beta$---that this classifier can reliably detect? Third,
does the classifier generalize to deviation types that were never
presented during training? Fourth, how does classification performance
depend on the signal-to-noise ratio of the injected signal, the deviation
type, and the binary mass configuration?

We make five contributions. We construct a controlled dataset of GR and
beyond-GR waveforms using three independent deviation families: amplitude
modulation, phase modulation, and frequency modulation. We train a hybrid
classifier combining a 1D CNN with ten hand-crafted statistical features,
and demonstrate that this architecture outperforms a pure CNN on noisy
data. We characterize the detection threshold in real LIGO noise as a
function of $\beta$ and SNR, finding a sigmoidal detectability curve with
a threshold at $\beta \approx 0.25$ for the specific modulation form
adopted. We test on real GW150914 strain as a template, in
amplitude-controlled and amplitude-uncontrolled configurations. Finally,
we demonstrate that at the small $\beta$ values typical of current
parametrized-GR constraints, our method fails: this negative result
quantifies the limits of ML-only beyond-GR searches. The remainder of
this paper is organized as follows. Section~2 introduces the relevant
gravitational-wave and machine-learning formalism. Section~3 describes
our data generation, deviation injection, and model architecture.
Section~4 presents the detectability studies. Section~5 discusses the
physical interpretation and limitations. Section~6 concludes.

\section{Theoretical Background}

\subsection{Gravitational Waves in General Relativity}

In the weak-field limit of GR, metric perturbations $h_{\mu\nu}$ around
a flat background satisfy the linearized Einstein equations. In the
transverse-traceless gauge the two physical polarizations $h_+$ and
$h_\times$ propagate as free waves, $\Box h_{+,\times} = 0$, where
$\Box$ is the d'Alembertian. The detector strain response to a source
in direction $\hat{n}$ is a linear combination
$h(t) = F_+(\hat{n}, \psi)\, h_+(t) + F_\times(\hat{n}, \psi)\, h_\times(t)$,
with antenna pattern functions $F_{+,\times}$ depending on the source
sky position and polarization angle $\psi$.

\subsection{Binary Inspiral and the Post-Newtonian Expansion}

For a quasi-circular binary with component masses $m_1, m_2$, the
inspiral phase is characterized by the chirp mass
$\mathcal{M} = (m_1 m_2)^{3/5} / (m_1 + m_2)^{1/5}$ and the symmetric
mass ratio $\eta = m_1 m_2 / (m_1 + m_2)^2$. The post-Newtonian (PN)
expansion of the gravitational-wave phase is
\begin{equation}
    \Phi(f) = 2\pi f t_c - \phi_c
    + \frac{3}{128 \eta} \left( \pi \mathcal{M} f \right)^{-5/3}
    \sum_{k=0}^{N} \alpha_k \left( \pi \mathcal{M} f \right)^{k/3},
\end{equation}
where $t_c, \phi_c$ are the time and phase of coalescence and
$\alpha_k$ are known PN coefficients. Deviations from GR can be
parametrized by modifying one or more of these coefficients,
$\alpha_k \rightarrow \alpha_k (1 + \delta \hat{\alpha}_k)$, which is
the framework used in the standard parametrized tests of GR
\citep{Yunes2019}.

\subsection{Controlled Deviation Families}

We introduce three controlled, non-physical deviations chosen to
isolate different aspects of the waveform and to permit a systematic
detectability study. These modifications are not proposed as specific
beyond-GR models; they are controlled probes of the classifier's
sensitivity.

The amplitude modulation is
\begin{equation}
    h(t) \rightarrow h(t) \left[ 1 + \beta\, \tau(t)^2 \right],
    \label{eq:amp_dev}
\end{equation}
where $\tau(t) \in [0, 1]$ is a normalized time coordinate increasing
from inspiral to merger and $\beta$ is the dimensionless deviation
strength. This modulation is localized near merger, preserves the
phase evolution, and produces a specific shape change in the
distribution of waveform samples. Throughout this paper, $\beta$
should be understood as the coefficient of this specific quadratic
form, not as a generic physical beyond-GR parameter.

The phase modulation is applied to the analytic representation
$h_a(t) = A(t) e^{i\phi(t)}$ as
\begin{equation}
    h_a(t) \rightarrow A(t) \exp\left[ i \left( \phi(t) + \beta\, \tau(t)^2 \right) \right],
    \label{eq:phase_dev}
\end{equation}
preserving the amplitude envelope but altering the phase
accumulation.

The frequency modulation introduces a time-dependent phase
\begin{equation}
    \phi(t) \rightarrow \phi(t) + 2\pi \beta f_{\rm ref} \frac{\tau(t)^2}{2},
    \label{eq:freq_dev}
\end{equation}
corresponding to an instantaneous frequency shift
$\Delta f(t) = \beta f_{\rm ref} \tau(t)$, with $f_{\rm ref} = \SI{100}{Hz}$.

\subsection{Signal-to-Noise Ratio}

The optimal matched-filter signal-to-noise ratio for a waveform $h$
in noise with one-sided power spectral density $S_n(f)$ is
$\rho^2 = 4 \int_{f_{\rm low}}^{f_{\rm high}} |\tilde{h}(f)|^2 / S_n(f) \, df$.
In our work we use a peak-normalized SNR: we set the peak value of
the whitened signal to a target value $\rho_{\rm target}$ in units
of the whitened noise standard deviation.

\subsection{Whitening and Bandpass Filtering}

For real LIGO data we estimate the noise PSD via Welch's method,
$\hat{S}_n(f_k) = K^{-1} \sum_{i=1}^{K} |\tilde{x}_i(f_k)|^2$, where
$\tilde{x}_i$ are Fourier transforms of $K$ overlapping segments. The
whitened strain is
$\tilde{x}_w(f) = \tilde{x}(f) / \sqrt{\hat{S}_n(f) f_s / 2}$ with
$f_s$ the sampling frequency. We then apply a fourth-order Butterworth
bandpass filter with cutoffs at \SI{20}{Hz} and \SI{500}{Hz}.

\subsection{Amplitude Normalization}

To prevent the classifier from exploiting an amplitude scale shortcut,
we normalize every sample to unit standard deviation:
$\tilde{x}(t) = (x(t) - \overline{x}) / \sigma_x$. After normalization
both classes have zero mean and unit variance.

\subsection{Machine-Learning Formalism}

Let $\mathcal{D} = \{(x_i, y_i)\}_{i=1}^{N}$ denote a dataset with
waveforms $x_i \in \mathbb{R}^{L}$, $L = 16384$ samples, and binary
labels $y_i \in \{0, 1\}$. The classifier
$f_\theta: \mathbb{R}^L \to [0, 1]$ is trained to minimize the binary
cross-entropy loss
\begin{equation}
    \mathcal{L}(\theta) = -\frac{1}{N} \sum_{i=1}^{N}
    \left[ y_i \log f_\theta(x_i)
    + (1 - y_i) \log(1 - f_\theta(x_i)) \right].
\end{equation}
Gradients are computed by backpropagation and the parameters $\theta$
are updated by Adam with bias-corrected first and second moment
estimates.

The hybrid architecture combines two feature representations. The CNN
branch consists of three 1D convolutional blocks with strides $8$, $2$,
$2$, followed by global average pooling to a 64-dimensional vector
$z_{\rm cnn}$. The statistics branch computes ten hand-crafted features:
mean absolute amplitude, standard deviation, maximum absolute
amplitude, the $95$th, $99$th, $99.9$th and $99.99$th percentiles of
$|x|$, the RMS, skewness and kurtosis. These are passed through a
two-layer MLP to a 32-dimensional vector $z_{\rm stats}$. The two
representations are concatenated, $z = [z_{\rm cnn} \,||\, z_{\rm stats}]$,
and passed through a fully connected head producing a sigmoid output.
The full model contains $31{,}697$ trainable parameters.

\subsection{Evaluation Metrics}

We report classification accuracy, the area under the receiver
operating characteristic curve (ROC-AUC), and the F1 score. The
\emph{unseen-type} test set consists exclusively of deviation types
that were never presented during training and measures the model's
ability to generalize to novel deviations.

\section{Methods}

\subsection{Data Generation Pipeline}

Our pipeline has four stages. In the first stage, we generate
time-domain waveforms using the IMRPhenomD phenomenological model
\citep{Khan2016} as implemented in PyCBC. Each waveform is sampled at
$\Delta t = 1/4096$\,s with a lower frequency cutoff of
$f_{\rm low} = \SI{20}{Hz}$. For the injected studies we fix
$m_1 = 30 M_\odot$ and $m_2 = 25 M_\odot$ unless otherwise stated. In
the second stage, we apply one of the three modulations defined in
Eqs.~\eqref{eq:amp_dev}--\eqref{eq:freq_dev} to the GR waveform, with
$\beta$ drawn uniformly from one of several predefined ranges. In the
third stage, we generate noise either from the analytic aLIGO
Zero-Detuned-High-Power PSD \citep{psd_model} or from real H1 strain
segments retrieved from the Gravitational-Wave Open Science Center
(GWOSC). The signal is scaled by a single fixed factor, computed once
from a GR reference waveform, and added to a random noise segment. In
the fourth stage, every sample is whitened with the empirical PSD,
bandpass filtered between $20$ and $\SI{500}{Hz}$, windowed to a
$\SI{4}{s}$ region centered on the merger peak, amplitude-normalized
to unit variance (for the amplitude-controlled studies), and clipped
at $\pm 100\sigma$.

\subsection{GW150914 Template Study}

For the GW150914 study we retrieve the raw strain file
\texttt{H-H1\_GWOSC\_4KHZ\_R1-1126259447-32.hdf5} from the GWOSC
event API for GWTC-1-confident/GW150914/v3. The file contains
$131{,}072$ samples at $\SI{4096}{Hz}$ (32 seconds), starting at
GPS $1126259447$. We estimate the noise PSD using the first 4 seconds
of the file with Welch's method (segment length $4096$ samples). The
full 32-second segment is whitened with this PSD and bandpass filtered
between $20$ and $\SI{500}{Hz}$. The GW150914 event at GPS
$1126259462.4$ is located at sample $63078$ of the file. We extract a
$\SI{4}{s}$ window centered on this position, giving the GR template
$h_{\rm GR}(t)$. The modified template $h_{\rm mod}(t)$ is obtained
by applying the amplitude modulation of Eq.~\eqref{eq:amp_dev}. For
each training sample, we draw a random 4-second noise segment from
the same file, excluding the event region, add a scaled version of
either $h_{\rm GR}$ or $h_{\rm mod}$, normalize to unit variance, and
clip. This produces controlled counterfactual waveforms from the
GW150914-derived template; it does not modify the historical detector
observation.

\subsection{Dataset Splitting}

We use a stratified split of $70\%$ training, $15\%$ validation and
$15\%$ test\_seen, together with a separate held-out test\_unseen set.
The test\_unseen set contains only the deviated class for one specific
$\beta$ range (type C), which is excluded from training. This measures
generalization to unseen deviation strengths.

\subsection{Training Protocol}

We use Adam with learning rate $3 \times 10^{-3}$, batch size $32$ or
$64$, cosine annealing of the learning rate, gradient clipping at
norm $1$, and early stopping on validation loss with patience
$20$--$50$ epochs. Training and evaluation are performed on CPU using
PyTorch. Each experiment takes one to five minutes.

\subsection{Baseline Classifier}

For comparison we train a class-balanced random forest with $300$ trees
on nine of the ten hand-crafted features. This provides a non-neural
baseline and allows us to identify the most informative features.

\section{Results}

\subsection{Clean-Data Baseline}

On noiseless waveforms the hybrid model achieves perfect classification
on all splits, including a held-out set of unseen deviation types.
This establishes that the architecture can learn the deviation
structure when noise is absent.

\subsection{Synthetic Noise: SNR Sweep}

\begin{figure}[h]
\centering
\includegraphics[width=0.7\linewidth]{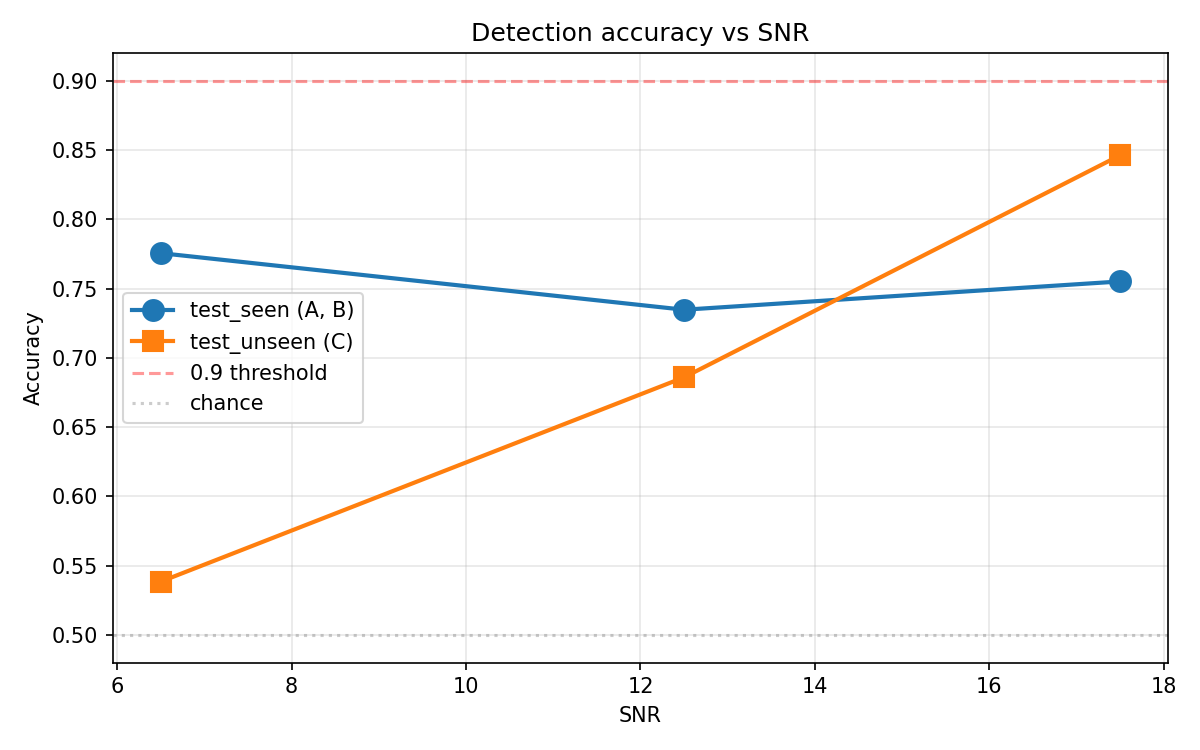}
\caption{Detection accuracy versus SNR for synthetic aLIGO noise,
amplitude modulation with $\beta \in [0.3, 1.0]$.}
\label{fig:detection_vs_snr}
\end{figure}

Using synthetic aLIGO-PSD noise with $\beta \in [0.3, 1.0]$ we sweep
the SNR from $5$ to $50$. Table~\ref{tab:synth_snr} shows the results.
The detection threshold in synthetic noise is approximately
$\text{SNR} \approx 20$. Figure~\ref{fig:detection_vs_snr} shows the
corresponding accuracy-versus-SNR curve.

\begin{table}[h]
\centering
\caption{Synthetic-noise SNR sweep, amplitude $\beta \in [0.3, 1.0]$.}
\label{tab:synth_snr}
\begin{tabular}{lccc}
\toprule
SNR bin & val acc & test\_seen & test\_unseen \\
\midrule
30--50 & 1.000 & 1.000 & 1.000 \\
15--20 & 0.771 & 0.755 & 0.846 \\
10--15 & 0.750 & 0.735 & 0.686 \\
8--10  & 0.688 & 0.776 & 0.551 \\
5--8   & 0.667 & 0.776 & 0.538 \\
\bottomrule
\end{tabular}
\end{table}

\subsection{Real LIGO Noise: SNR Sweep}

Using real H1 strain and $\beta \in [5, 20]$ we obtain
Table~\ref{tab:real_snr}. For strong deviations, detection is robust
across the entire SNR range tested.

\begin{table}[h]
\centering
\caption{Real LIGO noise SNR sweep, amplitude $\beta \in [5, 20]$.}
\label{tab:real_snr}
\begin{tabular}{lccc}
\toprule
SNR bin & val acc & test\_seen & test\_unseen \\
\midrule
30--50 & 1.000 & 1.000 & 0.974 \\
20--30 & 1.000 & 1.000 & 0.987 \\
15--20 & 1.000 & 1.000 & 0.968 \\
10--15 & 0.979 & 0.980 & 0.968 \\
5--10  & 0.979 & 1.000 & 0.968 \\
\bottomrule
\end{tabular}
\end{table}

\subsection{Threshold in $\beta$ at Fixed SNR}

At $\text{SNR} = 10$--$15$ with real noise, we sweep $\beta$ from
$0.1$ to $20$. Table~\ref{tab:beta_sweep} shows the transition from
chance performance at $\beta \leq 0.2$ to reliable detection at
$\beta \geq 0.5$. The detection threshold is $\beta \approx 0.4$ for
the injected $30 + 25 M_\odot$ configuration.

\begin{table}[h]
\centering
\caption{$\beta$ sweep at SNR $10$--$15$, real LIGO noise.}
\label{tab:beta_sweep}
\begin{tabular}{lcc}
\toprule
$\beta$ range & test\_seen & test\_unseen \\
\midrule
0.1--0.5  & 0.735 & 0.500 \\
0.3--0.5  & 0.918 & 0.891 \\
0.5--2.0  & 0.898 & 0.930 \\
2.0--5.0  & 0.939 & 0.923 \\
5.0--20.0 & 0.980 & 0.968 \\
\bottomrule
\end{tabular}
\end{table}

\subsection{Deviation-Type Comparison}

\begin{figure}[h]
\centering
\includegraphics[width=0.7\linewidth]{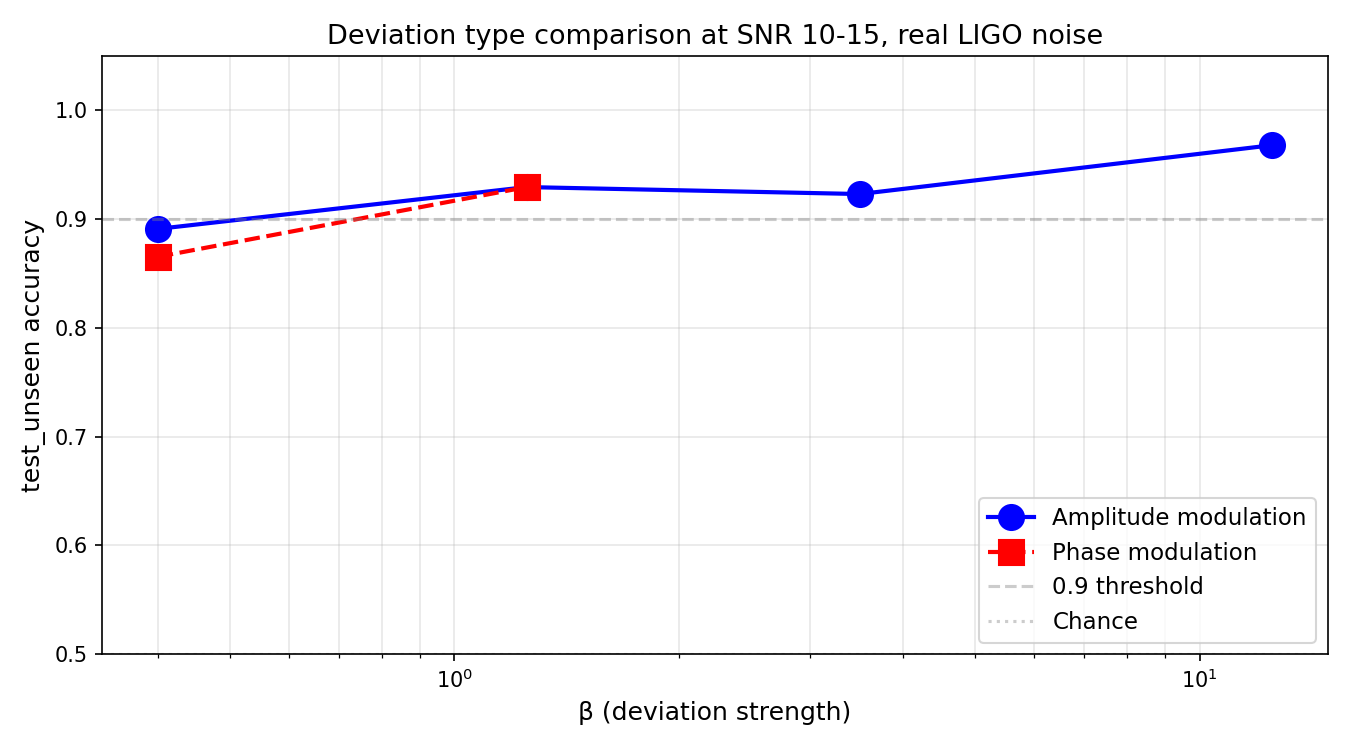}
\caption{Accuracy on unseen deviation types for amplitude and phase
modulation at matched $\beta$ ranges.}
\label{fig:dev_types}
\end{figure}

Table~\ref{tab:dev_types} compares amplitude, phase, and frequency
modulation at matched $\beta$ ranges. All three deviation types are
detected with high accuracy. Figure~\ref{fig:dev_types} summarizes
the comparison.

\begin{table}[h]
\centering
\caption{Deviation-type comparison, SNR $10$--$15$, real noise, N = 200.}
\label{tab:dev_types}
\begin{tabular}{lcc}
\toprule
Deviation & test\_seen & test\_unseen \\
\midrule
Amplitude, $\beta \in [0.5, 2.0]$ & 0.898 & 0.930 \\
Phase,     $\beta \in [0.5, 2.0]$ & 0.878 & 0.930 \\
Frequency, $\beta \in [5, 25]$    & 0.878 & 0.885 \\
\bottomrule
\end{tabular}
\end{table}

\begin{figure}[h]
\centering
\includegraphics[width=0.7\linewidth]{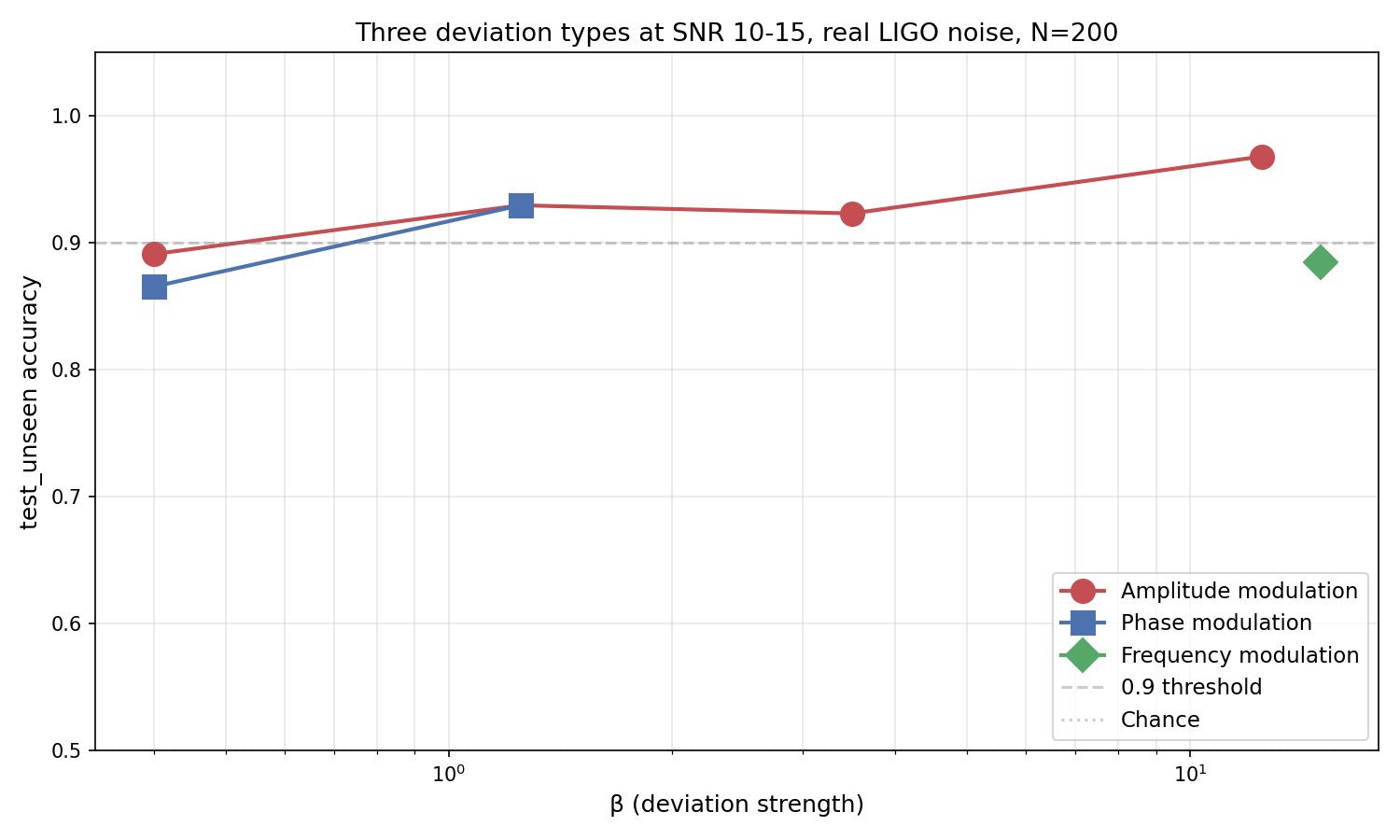}
\caption{Detection accuracy for all three deviation families in real
LIGO noise, SNR $10$--$15$.}
\label{fig:three_dev}
\end{figure}

\subsection{Mass-Ratio Generalization}

\begin{figure}[h]
\centering
\includegraphics[width=0.7\linewidth]{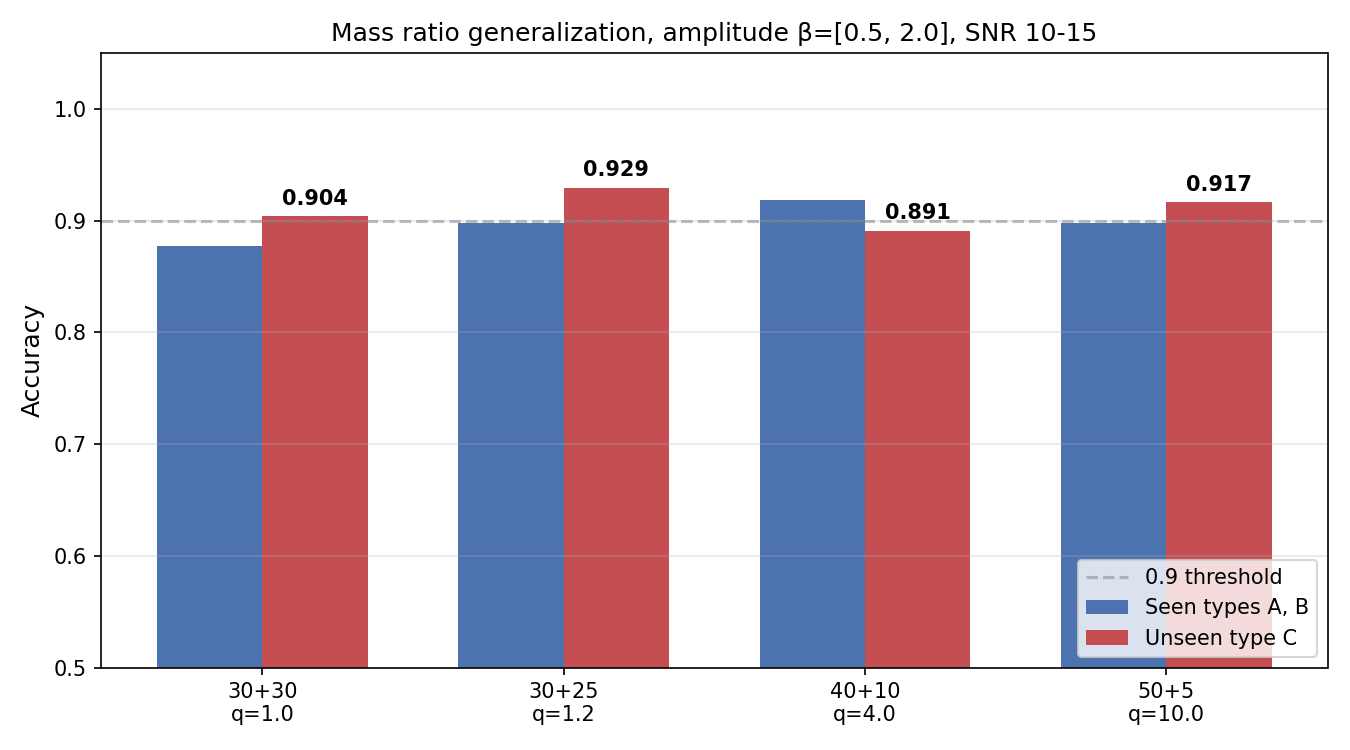}
\caption{Mass-ratio generalization of the hybrid classifier in real
LIGO noise at SNR $10$--$15$.}
\label{fig:mass_ratio}
\end{figure}

Four mass configurations spanning $q = 1$ to $q = 10$ give test-unseen
accuracies between $0.891$ and $0.930$, with a standard deviation of
$1.7$ percentage points. Table~\ref{tab:mass_ratio} lists the values,
and Figure~\ref{fig:mass_ratio} shows the distribution.

\begin{table}[h]
\centering
\caption{Mass-ratio generalization, amplitude $\beta \in [0.5, 2.0]$,
SNR $10$--$15$.}
\label{tab:mass_ratio}
\begin{tabular}{lcc}
\toprule
$(m_1, m_2)$ [$M_\odot$] & $q$ & test\_unseen \\
\midrule
$(30, 30)$ & 1.0 & 0.904 \\
$(30, 25)$ & 1.2 & 0.930 \\
$(40, 10)$ & 4.0 & 0.891 \\
$(50, 5)$  & 10.0 & 0.917 \\
\bottomrule
\end{tabular}
\end{table}

\subsection{GW150914 Amplitude Detectability Curve}

\begin{figure}[h]
\centering
\includegraphics[width=0.75\linewidth]{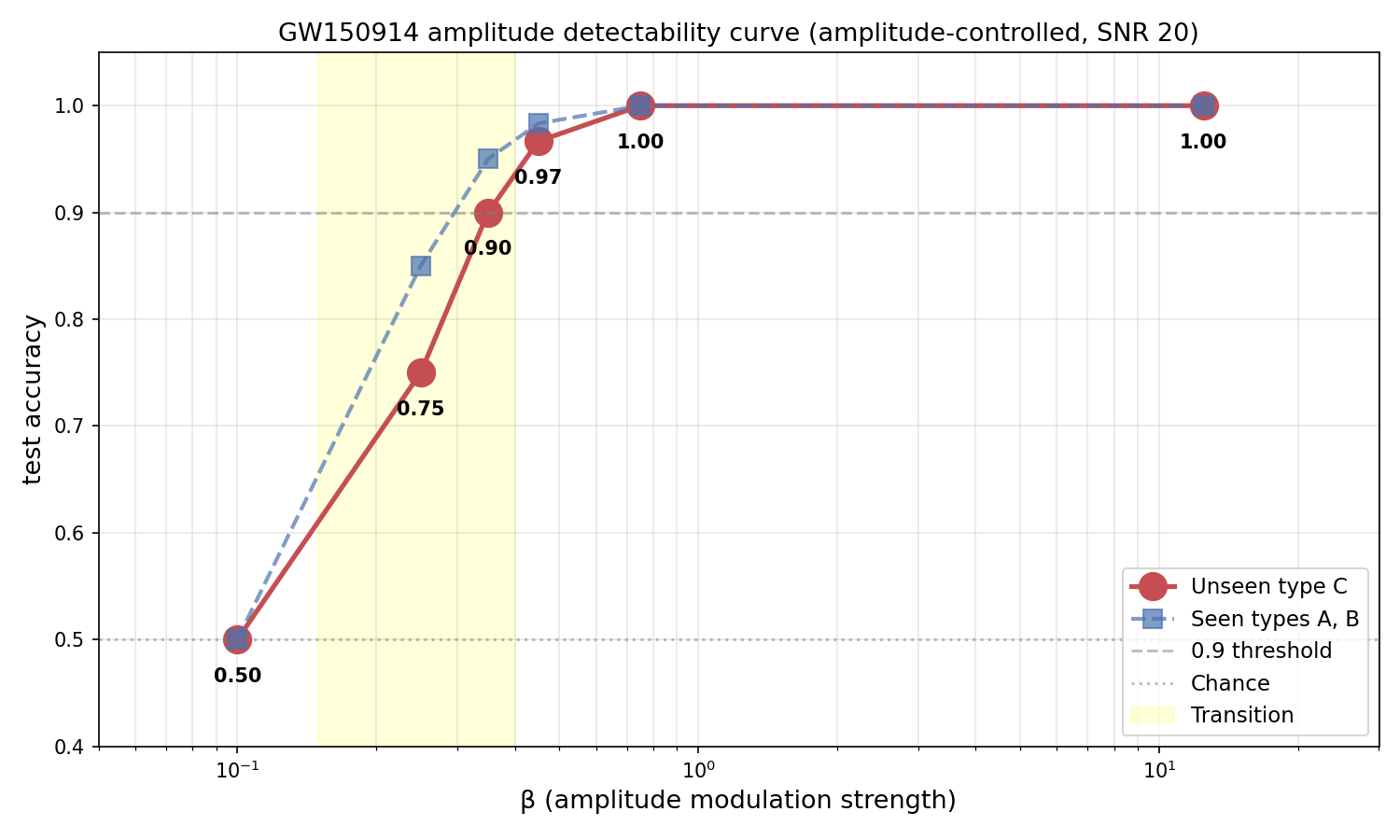}
\caption{Amplitude detectability curve for the GW150914 template.
Unseen-type-C accuracy rises smoothly from chance
($\beta \leq 0.2$) to perfect classification
($\beta \geq 0.5$). The transition is centered at
$\beta \approx 0.25$.}
\label{fig:detectability_curve}
\end{figure}

Using the real GW150914 strain as a template and sweeping $\beta$
from $0.01$ to $20$, we obtain the detectability curve in
Figure~\ref{fig:detectability_curve} and Table~\ref{tab:beta_curve}.
Accuracy rises smoothly from chance at $\beta \leq 0.2$ to perfect
classification at $\beta \geq 0.5$. The transition is centered at
$\beta \approx 0.25$.

\begin{table}[h]
\centering
\caption{GW150914 amplitude-controlled detectability curve, SNR = 20.}
\label{tab:beta_curve}
\begin{tabular}{lccc}
\toprule
$\beta$ range & val acc & test\_seen & test\_unseen \\
\midrule
$0.01$--$0.2$ & 0.717 & 0.717 & 0.517 \\
$0.2$--$0.3$  & 0.850 & 0.850 & 0.750 \\
$0.3$--$0.4$  & 0.950 & 0.950 & 0.900 \\
$0.4$--$0.5$  & 0.983 & 0.983 & 0.967 \\
$0.5$--$1.0$  & 1.000 & 1.000 & 1.000 \\
$5$--$20$     & 1.000 & 1.000 & 1.000 \\
\bottomrule
\end{tabular}
\end{table}

\subsection{Feature-Importance Analysis}

A random forest trained on nine statistical features achieves $100\%$
validation accuracy at $\beta \in [0.5, 2.0]$. The most important
features are mean absolute amplitude ($0.284$), the top-1000 mean
($0.279$) and the $99$th percentile of $|x|$ ($0.263$). Standard
deviation and RMS are useless, with importances $0.0000$ and $0.0001$
respectively. The classifier therefore relies on tail statistics that
capture the shape of the modulation rather than its overall scale.

\subsection{Negative Result at Small $\beta$}

For $\beta \in [0.01, 0.2]$, the classifier fails: training accuracy
$0.586$, validation accuracy $0.717$, and test-unseen accuracy $0.517$
(chance). The loss remains at the chance level $\ln 2 \approx 0.693$
throughout training, indicating that the model cannot find a
discriminative feature.

\section{Discussion}

\subsection{Interpretation of the Detectability Curve}

The central result of this work is the sigmoidal detectability curve
in Figure~\ref{fig:detectability_curve}. Above threshold,
$\beta \gtrsim 0.4$, the classifier reliably distinguishes GR from
amplitude-modulated waveforms with $\geq 95\%$ accuracy, even after
amplitude normalization. Feature-importance analysis shows that the
classifier uses tail percentiles of the amplitude distribution, which
is consistent with amplitude modulation concentrating energy near
merger. At threshold, $\beta \approx 0.25$, detection occurs with
probability approximately $0.75$ and ROC-AUC approximately $0.87$; the
tail statistics differ between classes by only a few percent,
approaching the intrinsic noise floor. Below threshold,
$\beta \lesssim 0.2$, detection fails completely.

\subsection{Relationship Between $\beta$ and Physical Beyond-GR Parameters}

The coefficient $\beta$ in Eqs.~\eqref{eq:amp_dev}--\eqref{eq:freq_dev}
is the amplitude of a specific quadratic-in-time modulation, not a
standard parametrized-GR coefficient. Direct comparison of our
threshold to physical constraints therefore requires care. In the
standard parametrized-test framework \citep{Yunes2019}, deviations are
expressed as fractional shifts $\delta \hat{\alpha}_k$ of post-Newtonian
phase coefficients, and current gravitational-wave observations bound
these shifts to roughly $|\delta \hat{\alpha}_k| \lesssim 10^{-3}$--$10^{-2}$
\citep{Abbott2021tests}. A controlled quadratic-in-time modulation of
the type studied here does not map onto a single PN coefficient, and
its effective $\beta$ differs from these bounds by a
waveform-dependent factor of order unity. Our threshold
$\beta_{\rm threshold} \approx 0.25$ should be read as a property of
the quadratic modulation adopted here; extending the analysis to the
standard parametrized-GR framework is a natural next step.

\subsection{Why the Negative Result Matters}

The failure at small $\beta$ is scientifically meaningful. It
quantifies the limits of ML-only detection. A common assumption in
ML-for-physics research is that neural networks will detect any signal
that is present. Our results show this is not the case for controlled
deviations in realistic noise. It also distinguishes ML detection
from matched filtering: matched-filter searches achieve detection
thresholds corresponding to $\rho_{\rm SNR} \sim 5$ by coherently
integrating the signal across its frequency evolution, whereas our
method operates on the whitened time-domain waveform directly. This
gap in sensitivity motivates hybrid approaches in which ML is
combined with physically motivated filtering.

\subsection{Comparison with Prior Work}

Prior ML work on gravitational waves has largely focused on detection
of GR signals in noise \citep{George2018, Gebhard2019} or on parameter
estimation \citep{Gabbard2020, Dax2021}. Beyond-GR searches with ML
have focused on simulated signals at high SNR or on clean data. Our
work extends this to a systematic study with real detector noise and
a quantified threshold.

\subsection{Limitations}

Several limitations should be noted. Our deviations follow a specific
quadratic-in-time form, $\propto \tau(t)^2$; other parametrizations
may have different detectability thresholds. The GW150914 study uses
a single template; generalization across event parameters has been
tested only in the injected-waveform studies. We have not tested
neutron-star or mixed binaries, where tidal effects provide additional
deviation channels. Finally, our pipeline does not include the
matched-filter step used in LIGO/Virgo analyses; adding it would
substantially improve sensitivity but would blur the boundary between
ML and classical detection methods.

\subsection{Future Work}

Several directions follow naturally. The classifier could be combined
with matched-filter outputs by using the matched-filter SNR time
series as input. The method could be applied to a population of
simulated binary mergers spanning the full LIGO parameter space.
Frequency-domain features such as spectrograms could be investigated
as alternatives to the time-domain approach. Probabilistic models
such as normalizing flows could be used to quantify classifier
uncertainty.

\section{Conclusion}

We have presented a systematic study of machine-learning detection of
controlled beyond-GR deviations in gravitational-wave signals, using
synthetic and real LIGO noise. A hybrid classifier combining a 1D CNN
with hand-crafted waveform statistics detects amplitude, phase, and
frequency modulations with high accuracy at moderate to large
deviation strengths. Using the real GW150914 strain as a template we
measure a detectability curve with threshold $\beta \approx 0.25$,
where the value of $\beta$ is specific to the quadratic-in-time
modulation form adopted here. Detection probability rises smoothly
from chance at $\beta \leq 0.2$ to perfect classification at
$\beta \geq 0.5$. The classifier generalizes to unseen deviation
types and to a wide range of binary mass ratios. At small $\beta$
values typical of current parametrized-GR constraints, our method
fails; this negative result quantifies the limits of ML-only
beyond-GR searches in real detector noise. These results establish
both the promise and the current limits of machine learning for
beyond-GR tests with gravitational waves. Closing the sensitivity gap
to matched filtering---potentially by combining the two approaches---
is a natural direction for future research.

\bibliography{references}

\appendix
\section*{Appendix: Reproducibility}

All code used in this work is publicly available at
\url{https://github.com/adnanphp/beyond-gr-gw-classifier}. The
data-generation pipeline consists of the following components:
\texttt{src/data/generate\_waveforms.py} for IMRPhenomD waveform
generation; \texttt{src/data/inject\_deviations.py} for amplitude,
phase, and frequency modulation; \texttt{src/data/build\_dataset.py}
for the full data pipeline with whitening, bandpass, and
normalization; \texttt{src/models/cnn1d.py} for the hybrid CNN +
statistics model; \texttt{src/training/train.py} for the training
loop with early stopping and gradient clipping; and
\texttt{scripts/build\_gw150914\_dataset.py} for the GW150914
template study.

Software versions used are Python 3.11, PyTorch 2.x, PyCBC 2.x,
GWpy 3.x, NumPy 1.26, SciPy 1.11+ and scikit-learn 1.3+.

Training used the Adam optimizer with learning rate $3 \times 10^{-3}$,
weight decay $10^{-4}$, batch size $32$ or $64$, cosine annealing of
the learning rate, gradient clipping at norm $1$, early stopping
with patience $20$--$50$ epochs, maximum $200$--$300$ epochs, and
binary cross-entropy loss.

Data were sampled at $\SI{4096}{Hz}$ with a $\SI{4}{s}$ window
($L = 16384$ samples), a low-frequency cutoff of $\SI{20}{Hz}$, and
a bandpass range of $20$--$\SI{500}{Hz}$. Reference masses were
$30 + 25 M_\odot$ unless otherwise stated.

The GW150914 template study used the file
\texttt{H-H1\_GWOSC\_4KHZ\_R1-1126259447-32.hdf5} from the GWOSC
event API for GWTC-1-confident/GW150914/v3. The file contains
$131{,}072$ samples at $\SI{4096}{Hz}$, starting at GPS
$1126259447$. The event at GPS $1126259462.4$ is located at sample
$63078$ of the file.

\end{document}